\documentclass[twocolumn,twoside,preprintnumbers,amsmath,amssymb,showkeys]{revtex4}
\usepackage{epsfig}
\usepackage{graphicx}
\usepackage{fancyhdr}

\usepackage{pslatex}

\begin{document}

\title{Event-by-event fluctuations of transverse-momentum and multiparticle clusters in
relativistic heavy-ion collisions \footnote{Talk presented by WF at the XXXVI International Symposium on Multiparticle Dynamics, Sept. 2 - 8, 2006, Paraty, Brazil}}

\author{W. Florkowski$^{1,2}$, W. Broniowski$^{1,2}$, B. Hiller$^3$, and P. Bo\.zek$^{2,4}$}

\affiliation{$^1$ Institute of Physics, \'Swi\c{e}tokrzyska Academy,
ul.~\'Swi\c{e}tokrzyska 15, PL-25406~Kielce, Poland \\
         $^2$ The H. Niewodnicza\'nski Institute of Nuclear Physics, 
Polish Academy of Sciences, PL-31342 Krak\'ow, Poland \\
         $^3$ Centro de F{\'{i}}sica Te\'orica, Departamento de F{\'{i}}sica, 
University of Coimbra, P-3004-516 Portugal\\
        $^4$ Rzesz\'ow University, PL-35959~Poland}


\begin{abstract}
We analyze the event-by-event fluctuations of mean transverse-momentum measured recently by the PHENIX and STAR Collaborations at RHIC. We argue that the observed scaling of strength of dynamical fluctuations with the inverse number of particles can be naturally 
explained by formation of multiparticle clusters. 


\end{abstract}
\maketitle

\thispagestyle{fancy}

\setcounter{page}{1}

\section{Introduction}

This paper is based on our recent analysis \cite{Broniowski:2005ae} of the event-by-event fluctuations of the mean transverse-momentum measured by PHENIX \cite{Adcox:2002pa} and STAR \cite{Adams:2003uw,Adams:2005ka} Collaborations at RHIC. The discussed new data has extended our knowledge of the phenomena studied intensely in recent years 
\cite{Gazdzicki:1992ri,Stodolsky:1995ds,Shuryak:1997yj,%
Mrowczynski:1997kz,Stephanov:1999zu,Voloshin:1999yf,Appelshauser:1999ft,Korus:2001au,%
Baym:1999up,Bialas:1999tv,Asakawa:2000wh,Heiselberg:2000fk,%
Pruneau:2002yf,Jeon:2003gk,Gavin:2003cb,Abdel-Aziz:2005wc}. One of the central issues 
in this field is whether the $p_T$ fluctuations at intermediate momenta result from jets \cite{Adler:2003xq,Liu:2003jf}. In this paper we argue that the observed scaling of strength of dynamical fluctuations may be naturally explained by formation of multiparticle clusters. From our point of view, the explanation of the data in terms of quenched jets represents only one possible option, other physical mechanisms leading to clustering also explain the data. 

Our reasoning is based on the following three observations: $i$)~the mean and the variance of the inclusive momentum distribution are practically constant at low centrality parameters $(c=0-30\%)$, 
\begin{eqnarray}
\langle M \rangle = \hbox{const.}\,, \;\;\;\;\;  \sigma_p = \hbox{const.}\,,
\label{const}
\end{eqnarray}
$ii$)~the variance of the average momenta for the mixed events is practically equal to the variance of the inclusive distribution divided by the average multiplicity, 
\begin{eqnarray}
\sigma_M^{\rm mix} \simeq \frac{\sigma_p}{\langle n \rangle },
\label{varmix}
\end{eqnarray}
finally, $iii$)~the difference of the experimental and mixed-event variances of mean $p_T$, denoted as $\sigma_{\rm dyn}^2$, scales as inverse multiplicity
\begin{eqnarray}
\sigma_{\rm dyn}^2 \equiv \sigma_M^2 - \sigma_M^{2, {\rm mix}} 
\sim \frac{1}{\langle n \rangle}. 
\label{scaledyn} 
\end{eqnarray}

As mentioned above, an explanation of this scaling can be provided by clustering in the expansion velocity. Moreover, the extracted value of $\sigma_{\rm dyn}$ is large at the expected scale provided by the variance of $p_T$, which indicates that the clusters should contain at least several particles in order to combinatorically enhance the magnitude to the observed level. This point is supported by the numerical calculations of the values of $\sigma_{\rm dyn}^2$ coming from the resonance decays and from thermal clusters in statistical models of heavy-ion collisions, which show that these values are very small \cite{Broniowski:2005ae,Kisiel:2005hn}.

\begin{table}[b]
\caption{\label{tab:data} 
Analysis  of the event-by-event fluctuations in the transverse momentum. Upper rows: 
the PHENIX experimental data at $\sqrt{s_{NN}} =$ 130~GeV  \cite{Adcox:2002pa}; 
middle rows: the mixed-event results; bottom rows: our way of looking at the data. 
One observes that to a good approximation
$ \sigma_M^{2, \rm mix} \simeq \sigma_p^2/\langle n \rangle$ and 
$ \sigma_{\rm dyn}^2 = (\sigma_M^2-\sigma_M^{2,{\rm mix}}) \sim 1/\langle n \rangle$. 
Except for the first two rows, all values are given in MeV. The errors in the last row 
reflect the unknown round-off errors in the data of the upper and middle parts.}
\begin{center}
\begin{tabular}{|l|c|c|c|c|}
\hline
centrality & 0-5\% & 0-10\% & 10-20\% & 20-30\% \\
\hline
$\langle n \rangle$ & 59.6 & 53.9 & 36.6 & 25.0 \\
$\sigma_n$ & 10.8 & 12.2 & 10.2 & 7.8 \\
$\langle M \rangle$ & 523 & 523 & 523 & 520 \\
$\sigma_p$ & 290 & 290 & 290 & 289 \\
$\sigma_M$ & 38.6 & 41.1 & 49.8 & 61.1 \\
\hline
$\langle M \rangle^{\rm mix}$ & 523 & 523 & 523 & 520 \\
$\sigma_M^{\rm mix}$ & 37.8 & 40.3 & 48.8 & 60.0 \\
\hline
$\sigma_p \sqrt{\frac{1}{\langle n \rangle}+\frac{\sigma_n^2}{\langle n \rangle^3}}$ & 38.2 & 40.5 & 49.8 & 60.8 \\
$\sigma_{\rm dyn} \sqrt{ \langle n \rangle}$ & $60.3 \pm 1.6$ & $59.2 \pm 1.5$ & 
$59.8 \pm 1.2$ & $57.7 \pm 1.1$ \\
\hline
\end{tabular}
\end{center}
\end{table}

\section{Simple way of doing statistical analysis}

In this Section we present elementary statistical considerations which form the basis of our treatment of the fluctuations. We use the notation where the letter $p$ denotes $|\vec{p}_T|$, $p_i$ is the value of $p$ for the $i$th particle, and $M = \sum_{i=1}^n p_i/n $ is the mean transverse momentum in an event of multiplicity $n$. Consider events of multiplicity $n$ and transverse momenta $p_1, p_2, \dots, p_n$. The multiplicity $n$ and the momenta are varying randomly from event to event.  The probability density of occurrence of a given momentum configuration is $P(n) \rho_n(p_1,\dots,p_n)$, where $P(n)$ is the multiplicity distribution and $\rho_n(p_1,\dots,p_n)$ is the conditional probability distribution of occurrence of $p_1,\dots,p_n$ in accepted events, provided we have the multiplicity $n$. Note that in general $\rho$ depends functionally on $n$, as indicated by the subscript. The normalization is
\begin{equation}
\label{norm}
\sum_{n} P(n)=1, \;\;\;\;
\int dp_1 \dots dp_n \rho_n(p_1,\dots,p_n)=1.
\end{equation}
The {\em marginal} probability densities are defined as
\begin{eqnarray}
\rho_n^{(n-k)} (p_1,\dots,p_{n-k}) \equiv \int dp_{n-k+1} \dots dp_n \rho_n(p_1,\dots,p_n),
\end{eqnarray} 
with $k=1,\dots,n-1$. These are also normalized to unity, as follows from Eq.~(\ref{norm}).
Since the number of arguments distinguishes the marginal distributions $\rho_n^{(n-k)}$,
in the following we drop the superscript $(n-k)$.  Further, we introduce the
following definitions
\begin{eqnarray}
\langle p \rangle_n &\equiv& \int dp \rho_n(p) p,  \\
{\rm var}_n(p) &\equiv& \int dp \rho_n(p) \left (p-\langle p \rangle_n \right )^2 , 
\nonumber \\
{\rm cov}_n(p_1,p_2) &\equiv& \int dp_1 dp_2 \left (p_1-\langle p \rangle_n \right ) 
\left (p_2-\langle p \rangle_n \right)\rho_n(p_1,p_2). \nonumber
\end{eqnarray}
The subscript $n$ indicates that the averaging is taken in samples of multiplicity $n$. 
We note in passing that the commonly used {\em inclusive} distributions are related 
to the marginal probability distributions in the following way:
\begin{eqnarray}
& & \hspace{-0.4cm} \rho_{\rm in}(x) \equiv 
\sum_n P(n) \int dp_1 \dots dp_n \sum_{i=1}^n \delta(x-p_i) 
\rho_n(p_1,\dots,p_n)  \nonumber \\
& & \hspace{-0.4cm} =  \sum_n n P(n) \rho_n(x),  \\
& & \hspace{-0.4cm} \rho_{\rm in}(x,y) \equiv \nonumber \\ 
& & \hspace{-0.4cm} \sum_n P(n) \int dp_1 \dots dp_n \sum_{i,j=1, j \neq i}^n 
\delta(x-p_i) \delta(y-p_j) \rho_n(p_1,\dots,p_n) \nonumber \\
& & \hspace{-0.4cm} = \sum_n n(n-1) P(n) \rho_n(x,y),
\label{incl}
\end{eqnarray}
which are normalized to $\langle n \rangle$ and $\langle n(n-1)\rangle$, respectively. For the variable $M= \sum_{i=1}^n p_i/n $ we find immediately
\begin{eqnarray}
\langle M \rangle &=& \sum_n P(n)\int dp_1 \dots dp_n M \rho_n(p_1,\dots,p_n)  \\
&=& \sum_n P(n) \langle p \rangle_n, \nonumber \\
\langle M^2 \rangle &=& \sum_n P(n)\int dp_1 \dots dp_n M^2 \rho_n(p_1,\dots,p_n) 
\nonumber \\
&=&
\sum_n \frac{P(n)}{n} \langle p^2 \rangle_n   \\
& & + \sum_n \frac{P(n)}{n^2} \left [ \sum_{i,j=1, j \neq i}^n {\rm cov}_n(p_i,p_j)  +  n(n-1)\langle p \rangle_n^2 \right ]. \nonumber
\end{eqnarray}
\begin{figure}[t]
\begin{center}
\includegraphics[width=7.5cm]{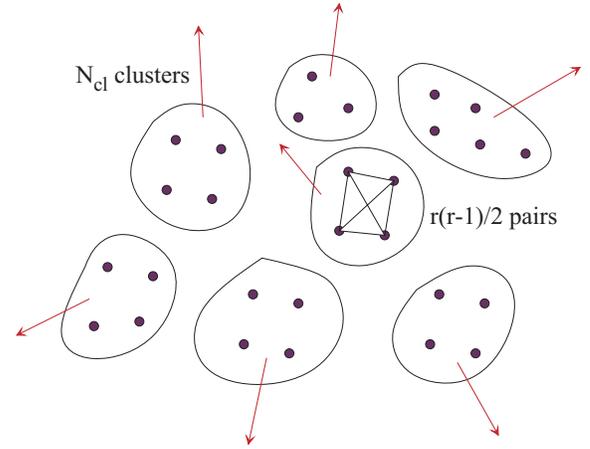}
\end{center}
\caption{The cluster picture of correlations. Particles are grouped in $N_{\rm cl}$ clusters, containing on the average $r$ particles. The particles within a cluster move at very similar collective velocities, indicated by arrows.}
\label{fig:clust}
\end{figure}

\section{Analysis of the PHENIX data}

At first we discuss the PHENIX measurement \cite{Adcox:2002pa} of the event-by-event fluctuations of the mean transverse momentum at $\sqrt{s_{NN}}=130$~GeV. We use the experimental fact that the variance of the momentum distribution and its mean
are independent of centrality (in the considered range), which allows us to replace the quantities $\langle p \rangle_n$ by $\langle M \rangle$ and $\langle p^2 \rangle_n - \langle p \rangle_n^2$ by $\sigma_p^2$ at the average multiplicity, denoted
as $\sigma^2_{p, \langle n \rangle}$. In this way we get
\begin{eqnarray}
\sigma_M^2 = \sigma_{p, \langle n \rangle}^2 \sum_n \frac{P(n)}{n} + 
\sum_n \frac{P(n)}{n^2} \left [ \sum_{i,j=1, j \neq i}^n   {\rm cov}_{n}(p_i,p_j) 
\right ]. 
\label{central}
\end{eqnarray}
In the mixed events, by construction, particles are not correlated, hence the covariance term 
in Eq.~(\ref{central}) vanishes and
\begin{eqnarray}
\sigma_M^{2,{\rm mix}} = \sigma_{p, \langle n \rangle}^2 \sum_n \frac{P(n)}{n} \simeq 
\sigma_{p, \langle n \rangle}^2 \left ( \frac{1}{ \langle n \rangle } 
+ \frac{\sigma_n^2}{\langle n \rangle^3} +
\dots \right ), 
\label{mixed}
\end{eqnarray}
where in the last equality we have used the fact that the distribution $P(n)$ 
is narrow and expanded $1/n=1/[\langle n \rangle +(n-\langle n \rangle)]$ to second order in $(n-\langle n \rangle)$. Comparison made in Table \ref{tab:data}, see the 8th and 9th rows,  shows that formula (\ref{mixed}) works at the $\mbox{1-2\%}$ level. In addition,
since $\sigma_{p, \langle n \rangle}$ is not altered by the event mixing
procedure, subtracting (\ref{mixed}) from (\ref{central}) yields
\begin{eqnarray}
\sigma_{\rm dyn}^2 &=&
\sum_n \frac{P(n)}{n^2}  \sum_{i,j=1, j \neq i}^n  {\rm cov}_{n}(p_i,p_j) \nonumber \\
&\simeq& \frac{1}{\langle n \rangle^2} \sum_{i,j=1, j \neq i}^{\langle n \rangle} 
 {\rm cov}_{\langle n \rangle}(p_i,p_j). 
\label{dyn}
\end{eqnarray}

\section{Cluster explanation}

The scaling of the dynamical fluctuations with $\langle n \rangle$, see the last row of Table \ref{tab:data} and Eq.  (\ref{scaledyn}), may be regarded as a constraint restricting the behavior of the covariance term. For instance, one can immediately conclude that if all particles were correlated to each other, $\sum_{i,j=1, j \neq i}^n {\rm cov}_{\langle n \rangle}(p_i,p_j)$ would be proportional to the number of pairs, and $\sigma_{\rm dyn}$ would not depend on $\langle n \rangle$ at large multiplicities. A natural explanation of the scaling (\ref{scaledyn}) comes from the cluster model, depicted in Fig.~\ref{fig:clust}. The system is assumed to have $N_{\rm cl}$ clusters, each containing (on the average) $r$ particles. Below we keep $r= {\rm const.}$ for simplicity (a more general case is discussed in Ref. \cite{Broniowski:2005ae}). The particles are correlated if and only if they belong to the same cluster, where the covariance per pair is $2\,{\rm cov}^\ast$. The number of correlated pairs within a cluster is $r(r-1)/2$. Some particles may be unclustered, hence the ratio of clustered to all particles is $\langle N_{\rm cl}\rangle r /\langle n \rangle = \alpha$. If all particles are clustered then $\alpha=1$. With these assumptions Eq.~(\ref{dyn}) becomes
\begin{eqnarray}
\sigma_{\rm dyn}^2 = \frac{\alpha(r-1)}{\langle n \rangle} {\rm cov}^\ast, 
\label{cluster}
\end{eqnarray}
which complies to the scaling (\ref{scaledyn}). An immediate conclusion here is that the ratio
$\alpha$ cannot depend on $\langle n \rangle$ (in the considered centrality range, $c=0-30\%$) in order for the scaling to hold. 

The question now is whether we can use the above results to draw conclusions on effects of jets (minijets), which have been proposed as a possible explanation of the experimental data even at the considered soft momenta \cite{Adler:2003xq}. Jets, when fragmenting, lead to clusters in the momentum space. The resulting full covariance from jets
is then  $N_{\rm cl, jet} j(j-1) {{\rm cov^j} \over \langle n \rangle^2}$, 
where $N_{\rm cl, jet}$ is the number of clusters originating from jets, $j$ is the average 
number of particles in the cluster, and $2\,{\rm cov^j}$ is the average covariance per pair.
The total number of particles produced from jets is $N_{\rm cl, jet} j$. 
On the other hand, the commonly accepted estimate of the dependence of $N_{\rm cl, jet} j$ on centrality 
is accounted for by the nuclear modification factor $R_{AA}$ multiplied by
the number of binary nucleon-nucleon collisons $N_{\rm bin}$. Since $R_{AA}$ depends on 
the ratio $\langle n \rangle / \langle n \rangle_{pp}$, where $\langle n \rangle_{pp}$ 
is the multiplicity in the proton-proton collisions, in a given $p_T$ bin one finds
\begin{eqnarray}
N_{\rm cl, jet}j \sim R_{AA} N_{\rm bin}= \frac{\langle n \rangle }{N_{\rm bin} 
\langle n \rangle_{pp} } N_{\rm bin} \sim \langle n \rangle, \label{jet}
\end{eqnarray} 
which complies to the scaling of Eq.~(\ref{cluster}).

\section{Conclusions}

The observed scaling of the dynamical fluctuations of mean transverse-momentum may be naturally explained by the cluster model. We emphasize that the scaling follows from the presence of clusters only, and is insensitive to the nature of their physical origin as long as one imposes the condition $N_{\rm cl} \sim \langle n \rangle$. 

The explanation of the observed data in terms of quenched jets agrees with the cluster picture. However, the explanation of the centrality dependence of the $p_T$ fluctuations in terms of jets based solely on Eq.~(\ref{jet}) is not decisive: any mechanism leading to clusters would do, for instance the thermal clusters discussed in Ref. \cite{Broniowski:2005ae} or mechanisms of Ref. \cite{Randrup}. Microscopic realistic estimates of the magnitude of ${\rm cov^j}$ and $j$ are necessary in that regard, including the interplay of the jets and medium. In short, the nature of clusters remains an open issue. For the current status of this program the reader is referred to \cite{Liu:2003jf,florence}.


\end{document}